\documentclass[preprint,1p]{elsarticle}

\usepackage{amssymb}
\usepackage{amsmath}
\usepackage{amsfonts}
\usepackage{algpseudocode}
\usepackage{array}
\usepackage[caption=false,font=normalsize,labelfont=sf,textfont=sf]{subfig}
\usepackage{listings}
\usepackage{textcomp}
\usepackage{stfloats}
\usepackage{url}
\usepackage{verbatim}
\usepackage{graphicx}
\usepackage{xcolor}
\usepackage[ruled,linesnumbered]{algorithm2e}

\journal{Computers \& Security}

\definecolor{verylightgray}{rgb}{.97,.97,.97}
\definecolor{applegreen}{rgb}{0.55, 0.71, 0.0}
\definecolor{iris}{rgb}{0.35, 0.31, 0.81}
\definecolor{richelectricblue}{rgb}{0.03, 0.57, 0.82}
\definecolor{shamrockgreen}{rgb}{0.0, 0.62, 0.38}
\definecolor{fulvous}{rgb}{0.86, 0.52, 0.0}
\definecolor{steelblue}{rgb}{0.27, 0.51, 0.71}
\definecolor{burgundy}{rgb}{0.5, 0.0, 0.13}
\definecolor{bittersweet}{rgb}{1.0, 0.44, 0.37}

\lstdefinelanguage{policyA}{
    basicstyle=\ttfamily,
    stepnumber=1,
    numbersep=8pt,
    showstringspaces=false,
    identifierstyle=\color{black},
    breaklines=true,
    backgroundcolor=\color{verylightgray},
    moredelim=**[is][\color{teal}]{\%}{\%},
}

\lstdefinelanguage{policyB}{
    basicstyle=\ttfamily,
    stepnumber=1,
    numbersep=8pt,
    showstringspaces=false,
    identifierstyle=\color{black},
    breaklines=true,
    backgroundcolor=\color{verylightgray},
    moredelim=**[is][\color{shamrockgreen}]{\%}{\%},
}

\lstdefinelanguage{policyC}{
    basicstyle=\ttfamily,
    stepnumber=1,
    numbersep=8pt,
    showstringspaces=false,
    identifierstyle=\color{black},
    breaklines=true,
    backgroundcolor=\color{verylightgray},
    moredelim=**[is][\color{shamrockgreen}]{\%}{\%},
    moredelim=**[is][\color{iris}]{∂}{∂},
}

\usepackage{xparse}
\newcounter{captionedlistingcounter}
\NewDocumentEnvironment{captionedlisting}{mmm}{
	\refstepcounter{captionedlistingcounter}
	{\centering\label{#1}}

}{

	\vspace{0.5\baselineskip}
	Listing \thecaptionedlistingcounter: #2
	\\
}

\usepackage[capitalize,nameinlink]{cleveref}
\crefname{line}{line}{lines}
\Crefname{Line}{Line}{Lines}
\crefname{captionedlistingcounter}{Listing}{Listings}
\Crefname{captionedlistingcounter}{Listing}{Listings}
\crefformat{footnote}{#2\footnotemark[#1]#3}

\begin{document}

\begin{frontmatter}



\title{Open Digital Rights Enforcement Framework (ODRE): from descriptive to enforceable policies}


\author[label1]{Andrea Cimmino} 

\author[label1]{Juan Cano-Benito}

\author[label1]{Raúl García-Castro}

\affiliation[label1]{
email={\{andreajesus.cimmino,juan.cano,r.garcia\}@upm.es}, 
organization={Ontology Engineering Group, Universidad Politécnica de Madrid},
            city={Madrid}, 
            country={Spain}}

\begin{abstract}
From centralised platforms to decentralised ecosystems, like Data Spaces, sharing data has become a paramount challenge. For this reason, the definition of data usage policies has become crucial in these domains, highlighting the necessity of effective policy enforcement mechanisms. The Open Digital Rights Language (ODRL) is a W3C standard ontology designed to describe data usage policies, however, it lacks built-in enforcement capabilities, limiting its practical application. This paper introduces the Open Digital Rights Enforcement (ODRE) framework, whose goal is to provide ODRL with enforcement capabilities. The ODRE framework proposes a novel approach to express ODRL policies that integrates the descriptive ontology terms of ODRL with other languages that allow behaviour specification, such as dynamic data handling or function evaluation. The framework includes an enforcement algorithm for ODRL policies and two open-source implementations in Python and Java. The ODRE framework is also designed to support future extensions of ODRL to specific domain scenarios. In addition, current limitations of ODRE, ODRL, and current challenges are reported. Finally, to demonstrate the enforcement capabilities of the implementations, their performance, and their extensibility features, several experiments have been carried out with positive results.


\end{abstract}




\begin{keyword}
Open Digital Rights Language \sep Privacy Policies \sep ODRL Enforcement


\end{keyword}

\end{frontmatter}


\section{Introduction}

In recent decades, proposals for sharing and consuming data from a set of wide-ranging domains among different actors have evolved; from large centralised data platforms to decentralised ones~\cite{10.1145/2295136.2295159}. Many of these proposals follow a policy-based approach~\cite{6984194}; such as European Data Spaces~\cite{akaichi2024interoperable}. On the one hand, they rely on a vocabulary that is used to unequivocally specify conditions and terms set by the data owner under which such data can be used by a third party, and, on the other hand, they rely on a software mechanism that must verify whether these conditions are met, that is, an enforcement procedure implementation. Some of these proposals are standards, such as XACML~\cite{standard2013extensible} or ODRL~\cite{odrl-standard}. However, not all of them provide enforcement, and some are only tailored to describe usage policies without the enforcement capabilities, like ODRL.

The Open Digital Rights Language (ODRL) is a W3C standard ontology that provides the vocabulary to describe policies in decentralised ecosystems, such as the Web, promoting data sovereignty. The ODRL vocabulary is widespread in numerous data sharing contexts as a mechanism to describe data licences~\cite{10.5555/2874359.2874375}, for example, extending ODRL to support GDPR privacy policies~\cite{de2019odrl}; which entails specifying certain abstract conditions to reuse data that cannot be enforced computationally. However, this standard goes beyond and allows describing more fine-grained enforceable conditions to use data (for example, data can be accessed only during the night), and actions are allowed to be performed on such data if conditions are met~\cite{10.1145/2660517.2660530}. Due to this reason, this standard is a candidate in many data-centric initiatives, such as European Data Spaces~\cite{eitel2021usage}.

However, the ODRL standard has only promoted a vocabulary to define its policies, but it lacks an enforcement specification. Therefore, policies cannot be evaluated to allow or deny data usage or invoke actions that must be taken if the constraints of these policies are met whether a third party actor requests the usage of a data resource~\cite{10.1007/978-3-030-89811-3_4}. Due to this reason, from the practical point of view, ODRL cannot be used to ensure the usage of data by third parties in the terms described by a data owner or to take the actions set by them~\cite{akaichi2022usage}. 

In addition, as ODRL policies are now endowed with this standard, they lack privacy in certain scenarios since they require stone-written conditions in its policies leading to privacy leak information, and, also, they lack a dynamic data handling mechanism which can also lead to privacy leaks. For instance, a policy based on geofencing that allows reading certain data to an actor if he/she is inside a place would show explicitly the geometry of such place (privacy leak) and will have no mechanism to handle the ever-changing GPS position of the actor (dynamic data) and, therefore, the GPS data will need to be continuously written also explicitly in the policy (privacy leak). 

In this article, the Open Digital Rights Enforcement (ODRE) framework is presented; which aims to extend the ODRL standard encompassing its ontology with enforcing features so policies described the ODRL vocabulary can be enforced. First, the article presents a novel approach on how these policies should be written so that they are not only descriptive, but also enforceable. This approach finds its roots in the way that behaviour and representation are addressed by the different types of grammar depending on the Chomsky classification~\cite{chomsky2014aspects} and how Model (behaviour) and View (representation) are separated in the template engines~\cite{parr2004enforcing}. The bottom line idea is to consider that ODRL policies are written with a descriptive language, and rely on adding one or more abstract languages to allow express behaviour, such as dynamic data values injection (data interpolation) or functions evaluation. To this end, a comprehensive analysis of the different types of policies that can be written and using different languages is presented, as well as some implementation languages that could be used.

Following this approach, the article then presents an implementation-agnostic algorithm, or procedure, that aligns with the ODRL standard vocabulary to enforce ODRL policies. This algorithm checks that the data usage restrictions described in the policies are met and then, if necessary, invokes related actions described in such policies. In addition, the algorithm is tailored with a mechanism for handling dynamic data (preventing privacy leaks) and counts with the necessary extension points to adapt ODRL policies to domain-specific scenarios. As part of the framework, a small extension of the ODRL has been developed and published~\footnote{\url{https://w3id.org/def/odre-time}} to showcase the extensibility feature of the enforcing algorithm.

For instance, the aforementioned geofencing policy requires firstly to extend the ODRL vocabulary to express the actor GPS position as a point and, also, to express the place as a geometry. Secondly, it requires a new functionality for the enforcement that is able to check whether the point is within the geometry. Finally, the policy must be written in a way that does not reveal the geometry or the GPS point, but it can be enforced with these values. Note that the enforcing algorithm and the ODRL vocabulary, or any extension, must be coupled. 

Finally, to prove the implementation-agnostic nature of the algorithm, the article reports two implementations published as open source; one based on Python\footnote{\url{https://pypi.org/project/pyodre/}} and the other on Java~\footnote{\url{https://github.com/ODRE-Framework/odre-java}}. These implementations have been tested with the same 24 unit tests each, and the time they required to enforce 24 different policies is reported. These tests and their related policies check different features of the implementations, such as their extensibility or implementation of the ODRL vocabulary.

The rest of this article is structured as follows. Section~\ref{sec:classification} introduces the ODRE framework approach to express ODRL policies so that they can become enforceable; Section~\ref{sec:algorithm} presents the ODRE framework algorithm to enforce policies with a running example and reports limitations of ODRE, ODRL and future challenges; Section~\ref{sec:experiments} presents two implementations of the ODRE enforce algorithm along with some experimental results; Section~\ref{sec:relatedwork} presents a survey of related proposals from the literature; finally, Section~\ref{sec:conclusions} presents a discussion about ODRL and its usage in practical scenarios and the conclusions of the article.

\section{Classification of ODRL enforceable policies}\label{sec:classification}

\begin{figure}[ht]
\centering
\includegraphics[scale=0.45]{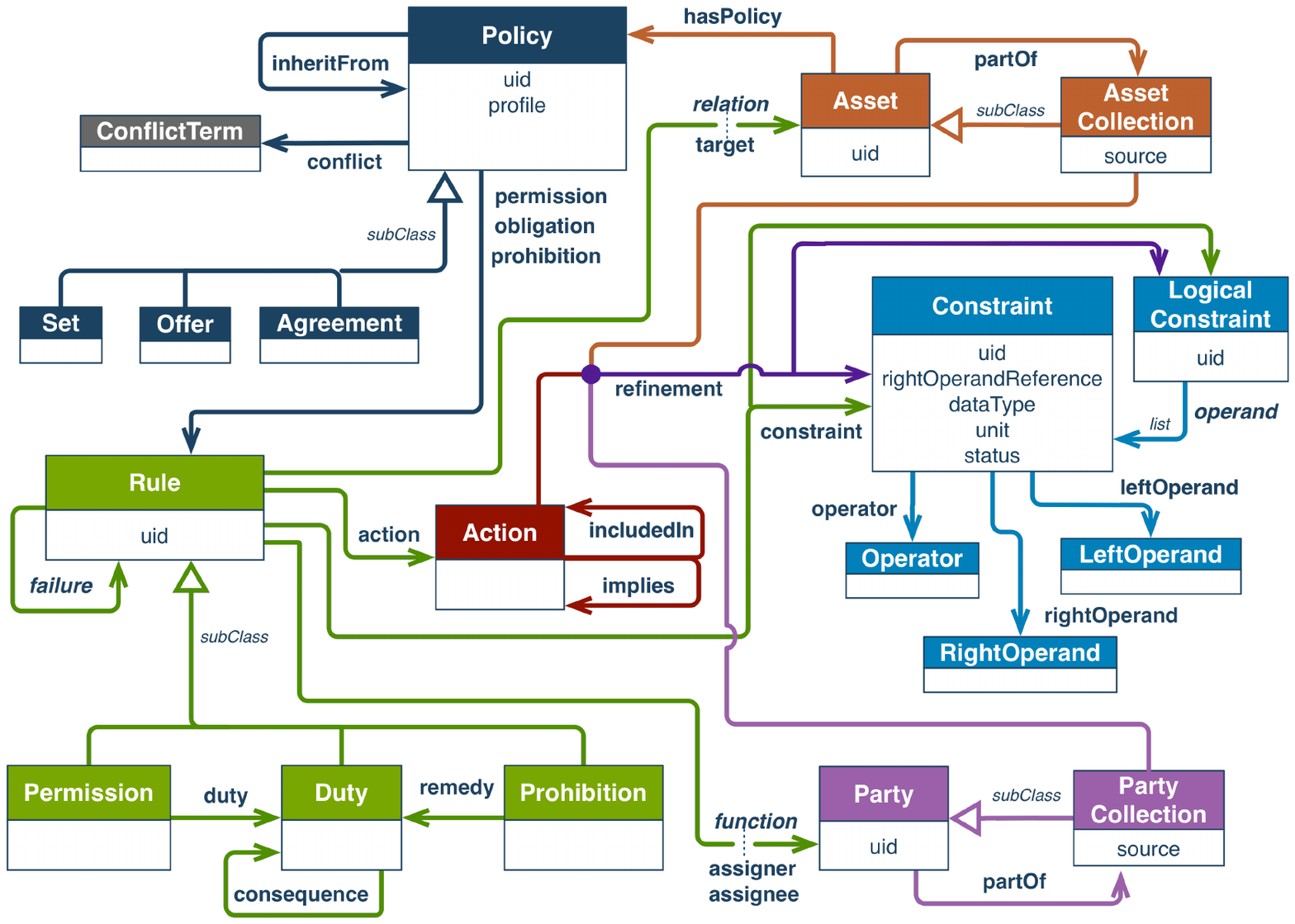}
\caption{ODRL vocabulary, extracted from~\cite{odrl-standard}}
\label{fig:odrlmodel}
\end{figure}

ODRL is an ontology that provides the vocabulary to describe or express usage policies following the model depicted in Figure~\ref{fig:odrlmodel}. From now on,  the prefix $odrl{:}$ for the ODRL ontology namespace\footnote{\url{http://www.w3.org/ns/odrl/2/}} is assumed. These policies may have a set of rules; which may denote permission, duty, or prohibition. Regardless of their type, rules consist of an action and a set of constraints that encode the conditions under which the action must be performed. The constraints have three main descriptive elements: an operator, a left operand, and a right operand. The operands must be either concepts defined in the ontology that represent a certain information that is not explicitly written in the policy (for example, $odrl{:}dateTime$\footnote{\url{https://www.w3.org/TR/odrl-vocab/#term-dateTime}} represents the current date with time), or data constants that have a value and a $xsd$ datatype\footnote{\url{https://www.w3.org/TR/xmlschema11-2/}} (for example, $\{ ``@value"{:} ``2018{-}01{-}01", ``@type"{:} ``xsd{:}date\-Time" \}$). Operators are always concepts defined in the ontology and represent predicates, i.e., functions that always output a Boolean result.

In order to enforce an ODRL policy, first its operands, which are ontology concepts, must be replaced in the policy by a data constant with the information they stand for. In the case of $odrl{:}dateTime$, a constant representing the actual date and time; which will look similar to the one shown above. Then, the operator must be computed taking the left and right operands as input. In the event the operator provides a positive result (true), the action of the policy must be taken; this may involve running some code, like for the operator, or notifying a third party (either a person or a software) who will perform such action. 

Note that ODRL policies are expressed in the W3C standard format RDF~\cite{manola2004rdf} which allows expressing concepts as triplets without any functionality or behaviour, and the ODRL ontology only narrows down the terms that a policy expressed in RDF can contain. As a result, it is necessary to wrap or transform the descriptive terms of the ODRL ontology into another language that can be interpreted and, therefore, that can replace operands for respective constants, evaluate operators, or run actions. In order to achieve this goal, the authors classify the ODRL policies into the following types depending on the ancillary language used.

The policies expressed in RDF with the ODRL ontology will be classified as $\emptyset{-}Level$ and the language used to write them as \textit{descriptive language} (in this case RDF with the terms from the ODRL ontology), and those expressed with an \textit{interpreted language} classified as $A{-}Level$. The result of interpreting an $A{-}Level$ policy is a $Usage$ $Decision$; that specifies a set of actions that should be performed either as part of the enforcement procedure or by notifying a third party who must perform them. These actions are present in the $Usage$ $Decision$ if their related constraints were enforced positively (true).\\

\textit{Note that $\emptyset{-}Level$ and $A{-}Level$ policies must show no differences in terms of lexicon, they both contain terms from the ODRL ontology, i.e., the descriptive language, as shown by~\cref{listing:levela}. However, the interpreted language couples certain terms (highlighted in~\cref{listing:levela}) from the descriptive language (i.e., ontology terms) with certain functionalities and, therefore, is able to interpret them and produce an enforcement result, i.e., Usage Decision. It is worth to mention that the terms that the interpreted language recognises must exist in the descriptive language, that is, the ontology.}

\begin{captionedlisting}{First Example}{A sample, $\emptyset{-}Level$ or $A{-}Level$, policy  that grants usage if datetime and media are greater or equals, respectively, to a certain value}
\centering
{\begin{lstlisting}[language=policyA,firstnumber=1]
{"@context": "...",
"@type": "Policy",
"uid": "https://upm.es/policy/1",
"permission": [{
   "target": "https://jsonplaceholder.typicode.com/users/1",
   "action": "read",
   "constraint": [{
      "leftOperand": "%dateTime%",
      "operator": "%gt%",
      "rightOperand": { "@value": "2018-01-01T00:40:30", "@type": "xsd:dateTime" }
   },{
      "leftOperand": "%media%",
      "operator": "%eq%",
      "rightOperand": { "@value": "online", "@type": "xsd:string" }
   }]
}]}
\end{lstlisting}}\label{listing:levela}
\end{captionedlisting}

Under certain circumstances, policies may need to contain certain terms that do not belong to the ODRL ontology and, instead, refer to data variables whose values shall be given by whoever is enforcing the policy; e.g., a third party an actor. These policies, classified as $B1{-}Level$, require an additional level of expressiveness provided by an additional language that wraps the $A{-}Level$ policies, that is, the \textit{interpolated language}. This language only allows to use terms that refer to data variables. \\

\textit{Note that $B1{-}Level$ policies, and following levels, introduce differences in terms of lexicon since they need to express terms outside the ODRL ontology and, therefore, these policies are written mixing RDF with the interpolated language. An example of a $B1{-}Level$ policy is shown by~\cref{listing:levelb1}. The interpolated language used in the policy, which is highlighted, is Freemarker\footnote{\url{https://freemarker.apache.org/docs/dgui_misc_alternativesyntax.html}}.}

\begin{captionedlisting}{B-Level Policy}{A sample $B1{-}Level$ policy that grants usage if data variable (\textit{requesterToken}) is equals to a certain value}
\centering
{\begin{lstlisting}[language=policyB,firstnumber=1]
{"@context": "...",
"@type": "Policy",
"uid": "https://upm.es/policy/1",
"permission": [{
   "target": "https://jsonplaceholder.typicode.com/users/1",
   "action": "read",
   "constraint": [{
      "leftOperand": { "@value": "%[=requestToken]%", "@type": "xsd:string" },
      "operator": "eq",
      "rightOperand": { "@value": "eyJhbGciO...", "@type": "xsd:string" }
    }]
 }]
}
\end{lstlisting}}\label{listing:levelb1}
\end{captionedlisting}

Most of the time, the interpolated language counts with more expressiveness than just terms to refer to data variables; providing additional ones for data flow control, iterative statements, or custom procedures that are defined relying only on this language; in this case, the policy is classified as $B2{-}Level$ and the language is known as \textit{templated language}, which is a richer or enhanced \textit{interpolated language}.\\

\textit{\cref{listing:levelb2} shows a  $B2{-}Level$ policy expressed with an extended version of the  \textit{templated language} from~\cref{listing:levelb1}. Note that this policy contains several procedures  concatenated after the procedure called now, which returns current time, and an variable creation using assign. Note that, in this language, control terms are preceded by the token [\# whereas data variables are preceded by the token [=.}

\begin{captionedlisting}{C-Level Policy}{A sample C-Level policy that grants usage if the current time is greater than 08:00 AM}
\centering
{\begin{lstlisting}[language=policyB,firstnumber=1]
{"@context": "...",
"@type": "Policy",
"uid": "https://upm.es/policy/1",
"permission": [{
   "target": "https://jsonplaceholder.typicode.com/users/1",
   "action": "read",
   "constraint": [{
 %[#assign timeVar=.now?time?iso("Europe/Rome")?replace('\\+.*','', 'r')]%
      "leftOperand":  { "@value": "%[=timeVar]%", "@type": "xsd:time" },
      "operator": "gt",
      "rightOperand": { "@value": "08:00:00", "@type": "xsd:time" }
    }]
 }]}
\end{lstlisting}}\label{listing:levelb2}
\end{captionedlisting}

Finally, policies may be wrapped into a higher abstract language that is able to code complex behaviour, these are known as $C{-}Level$ policies. This complex behaviour can be written using a \textit{coded language} and passed as a term of the \textit{templated language} or the \textit{coded language} can literally wrap the policy; which will be a data variable in the \textit{coded language}. In the former case, no new languages besides the \textit{templated language} will appear explicitly written in the policy. In the latter case, no reference to this complex behaviour will appear in the policy. It is recommended to avoid this last case since it may evolve in adding usage restrictions (i.e., constraints) in the code without appearing in the policy and, therefore, not using the ODRL ontology to express them. Conditions as constraints for the usage of a certain resource must always be expressed with the terms from the ODRL ontology.\\

\textit{\cref{listing:levelc} shows a sample $C{-Level}$ policy that relies on the complex function request (highlighted in blue) that performs a GET request to a remote API. This function is not implemented, and is not possible to be implemented, with a templated language as procedure. Instead it has to be written using a coded language, such as Python or Java, and passed as term to the templated language. After getting the data from the API, the policy filters its results and injects a fragment of data as a left operand of the policy.}

\begin{captionedlisting}{C-Level Policy}{A sample $C{-}Level$ policy that grants usage if temperature from external API is above 35º}
\centering
{\begin{lstlisting}[language=policyC,firstnumber=1]
%[#-- Data access --]
[#assign weather=%∂request("GET", https://api.open-meteo.com/v1/forecast?latitude=40.4050099&longitude=-3.839519&hourly=temperature_2m)∂%]
[#-- Data handling --]
[#assign tmp="ERROR"]
[#list weather.hourly.time as elem]
  [#if elem?contains("T"+currentTime) || elem?contains("T0"+currentTime)]
    [#assign tmp=weather.hourly.temperature_2m[elem?index]]
    [#break]
  [/#if]
[/#list]%

{"@context": "...",
"@type": "Set",
"@type": "Policy",
"uid": "https://upm.es/policy/1",
"permission": [{
   "target": "https://jsonplaceholder.typicode.com/users/1",
   "action": "read",
   "constraint": [{
      "leftOperand":  { "@value": "%[=tmp]%", "@type": "xsd:float"},
      "operator": "gt",
      "rightOperand":  { "@value": "35", "@type": "xsd:float" }
   } ]
}]}
\end{lstlisting}}\label{listing:levelc}
\end{captionedlisting}

\begin{figure}[ht]
\includegraphics[scale=0.87]{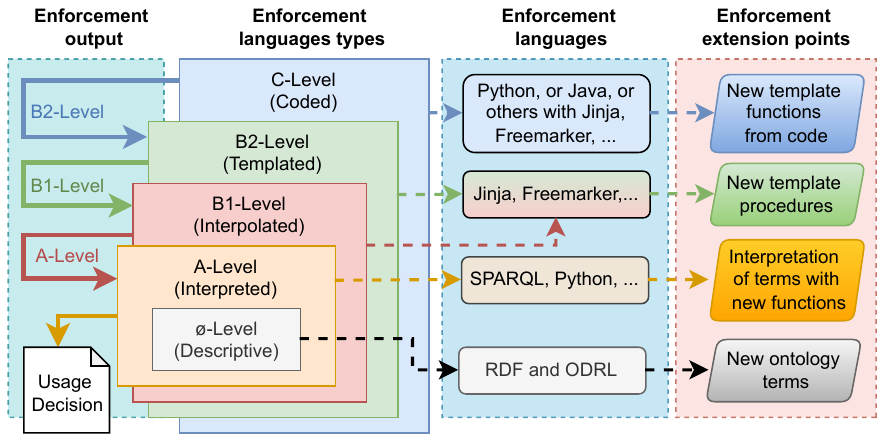}
\centering
\caption{ODRL enforcement hierarchy, languages, and extension points}
\label{fig:policiceshier}
\end{figure}

Figure~\ref{fig:policiceshier} shows the hierarchy between these languages and possible choices to implement them. Note that the enforcement occurs with the \textit{interpreted language}, which is the only one that cannot be omitted. Instead, the rest of the language layers on top of it, which are optional, are used to handle input dynamic data (\textit{interpolated and templated languages}) or to handle events outside the policy and use their data (\textit{coded languages}). In fact, when a $C{-}Policy$ is enforced, it must produce a $B2{-}Level$ or $B1{-}Level$ policy. Similarly, when $B2{-}Level$ is enforced, a $B1{-}Level$ policy is produced, and when $B1{-}Level$ is enforced a $A{-}Level$ policy. The only policies which result after the enforcement is a \textit{Usage Decision} are the $A{-}Level$ policies.

On the other hand, Figure~\ref{fig:policiceshier} shows several languages that can be used to implement these language layers. Note that although $A{-}Level$ policies do not introduce new terms outside the ODRL ontology lexicon, when they need to be enforced, its interpretable terms (operands, operators, and actions) must be extracted and expressed with the \textit{interpreted language} so that they can be evaluated and produce a result. In order to extract these terms, a JSON Path or a SPARQL query can be used. As \textit{interpreted language} many choices exists, for instance the SPARQL query language~\cite{sparql-standard} or even Python, as long as they implement functions and these can be coupled with the terms that appear in the policies. The next section delve into the details of the necessary algorithm to enforce these policies regardless of the implementation language, and two implementations are reported and tested. One relying on SPARQL as \textit{interpreted language}, Freemarker as \textit{templated language} and Java as \textit{coded language}. The other relies on Python as \textit{interpreted language}, Jinja as \textit{templated language} and Python as \textit{coded language}.

Note that the approach described in this section also provides different extension points, as shown in Figure~\ref{fig:policiceshier}. This feature is paramount since ODRL is general purpose-based, and specific domains will need to extend the ODRL policies to adapt to new usage conditions. To this end, first the \textit{descriptive language} must be extended, that is, the ODRL ontology, by adding new terms following the good and well-established ontological practices for it~\cite{suarez2015neon}. Then, it is necessary to implement new functions in the \textit{interpreted language} and couple them to the new terms created in the ontology. These two extension points allow practitioners to adapt ODRL to new domains. 

The \textit{templated language} can be extended with new procedures, which are functions defined with the same  \textit{templated language}. These are helpful for encoding repetitive and complex data handling processes into a single procedure that can be called several times; note that such procedures are defined within the policy. Finally, the \textit{coded language} can extend the functions that the \textit{templated language} can use. However, practitioners must be careful using the \textit{templated language} and \textit{coded language} as well as their extensions, as they can define usage restrictions using these languages instead of relying on ODRL ontology terms. The usage conditions must always be express uniquely with terms from the ODRL ontology, or extensions of it, and always handled with the \textit{interpreted language}. The \textit{interpolated, templated, and coded languages} are only meant, and must only be used, to handle dynamic data or hide stone-written data values to avoid leakage of private information.

\section{Enforcement Algorithm}\label{sec:algorithm}

This section presents the ODRE framework algorithm endowed for enforcing ODRL policies expressed the approach explained in the previous section. However, before delve into the details, several concepts used by the algorithm must be defined.\\

\textbf{Policy.} $P_{i}$ is defined as a policy written with the terms from the ODRL ontology (or any ontological extension of it) and which level is denoted by $i$ that takes as value the letter of the level it represents, for instance, an $A{-}Level$ policy is denoted as $P_A$. In the case the policy level is unknown, or does not matter (it could be either $A$, $B1$, $B2$, or $C$), the policy is denoted as $P_{*}$. \\

\textbf{Policy reduction.} Reducing a policy is the process of enforcing any policy which level is above $A{-}Level$ ($B1, B2,$ or $C$ levels) and producing as output an $A{-}Level$ policy. To this end, the procedure $reduce$ is defined, which receives as input a policy $P_{*}$, a set $M$ that contains data variables and their values, and a set $F$ that contains function variables and their implementations with a \textit{coded language}. Both sets can be empty in the case no dynamic data is handled or no functions are implemented with the \textit{coded language}. \\

\textbf{Policy filtering.} Three filtering operations are needed to enforce a given policy. To this end, the procedures $rules$, $constrais$, and $action$ are defined. The first, $rules$, takes as argument an $A{-}Level$ policy and returns a set $R_{id}$ containing all the ids of every rule written in the policy; each of one denoted as $r_{id}$. The second, $constrais$, takes as argument an $A{-}Level$ policy and the id of a rule it contains ($r_{id}$); produces as result a set $C_R$ containing tuples of the form $(o, o_{L}, o_{R})$ where $o$ is an operator, $o_L$ and $o_R$ are its left and right operand respectively. The third, $action$, takes as argument an $A{-}Level$ policy and the id of a rule it contains ($r_{id}$); provides as result the action associated to that rule denoted by $a$. \\

\textbf{Policy transformation.} In order to enforce the constrains of a policy these have to be expressed into an \textit{interpreted language}. To this end, the $transformConstraints$ procedure is defined; which takes as argument the set $C_R$ that contains tuples of the form $(o, o_{L}, o_{R})$ being $o$ an operator, and  being $o_L$ and $o_R$ its left and right operand respectively. As a result,  $transformConstraints$ produces a policy expressed using an \textit{interpreted language} denoted by $P_I$.

In order to enforce actions, these must be translated into an \textit{interpreted language} as long as such language supports that action, i.e., it knows how to interpret it. To this end, the $supported$ procedure is defined; which takes as argument an action $a$ and returns a Boolean value indicating whether such action is supported by the  \textit{interpreted language}. Also, the procedure $transformAction$ is defined that takes as input the action $a$ and returns such action expressed with the \textit{interpreted language}, $A_I$ so it can be evaluated.
\\

\textbf{Policy Evaluation.} A policy expressed using an \textit{interpreted language}, denoted by $P_I$, can be evaluated producing a Boolean value as a result. The procedure $evaluate$ is defined so that it receives as argument a policy $P_I$  written with the \textit{interpreted language}, and outputs a value $d$ that can be true or false.\\

\textbf{Perform Action.} An action expressed using an \textit{interpreted language}, denoted by $A_I$, can be evaluated (a.k.a performed) producing a result. Note that only actions coupled with procedures in the \textit{interpreted language} can be evaluated. The procedure $perform$ is defined so that it receives as argument an action $A_I$  written with the \textit{interpreted language}, and outputs any value as a result of performing such action denoted by V. Note that the result may be an empty value. \\

\textbf{Usage Decision.} After an $A{-}Level$ policy is enforced, a \textit{Usage Decision} is provided. This concept, denoted by $D$, is a set of tuples filled during the evaluation of the different constraints written in the policy and denoted by $C_R$. The tuples of the \textit{Usage Decision} $D$ have as the first element the action $a$, the value of the second element may differ depending on one of the following scenarios. If the action $a$ is not supported by the $interpreted language$, i.e., $supports(a)$ is false, the second element of the tuple is filled with the same action $a$ so a third party is informed about the fact that the action is not performed by the algorithm. When the action $a$ is coupled with a procedure in the $interpreted language$, i.e., $supports(a)$ is true, then the tuple contains as second argument the result of performing such action. \\

Taking the previously defined concepts as granted,~\cref{alg:enforcement} presents the ODRE framework enforcement algorithm. The algorithm takes as input a policy $P_*$, a set of data variables and their values ($M$), and a set of function variables and their implementations ($F$). As a result, the algorithm produces a $Usage$ $Decision$ set ($D$).

\SetKwComment{Comment}{/* }{ */}

\begin{algorithm}[hbt!]
\caption{ODRE Enforcement algorithm}\label{alg:enforcement}
\KwData{$P_{*},  M, F $}
\KwResult{$D$}

$D \gets \emptyset$\;
$P_{A} \gets reduce(P_{*}, M, F)$\;
$R_{id} \gets rules(P_A)$\;
\For{$r_{id} \in R_{id}$}{
    $C_R \gets constraints(P_A, r_{id})$\;
    $P_I \gets transformConstraints(C_R)$\;
    $d \gets evaluate(P_I)$\;
    \If{d}{
        $a \gets action(P_{A}, r_{id})$\;
        
        \If{$supported(a)$}{
            $A_I \gets transformAction(a)$\;
            $V \gets perform(A_I)$\;
            $D \gets D \cup \{ (a, V) \}$\;
        }\Else{
            $D \gets D \cup \{ (a, a) \}$\;
        }
      
    }
}
\end{algorithm}

The algorithm first reduces the input policy $P_*$ to an $A{-}Level$ policy (line 2). To this end, it relies on the $reduce$ procedure which, on the one hand, replaces the data and function variables written in the policy with either the \textit{interpolated} or \textit{templated} language with their values encoded in the sets $M$ or $F$, respectively. On the other hand, $reduce$ enforces the policy that, as a result, obtains a $A{-}Level$ one. Note that if the input is already a $A{-}Level$ policy, $reduce$ has no effect since it only works on the \textit{interpolated} or \textit{templated} languages.

Then, the algorithm extracts and iterates over the rules ids ($R_{id}$) written in the reduced policy $P_A$ (lines 3-4). For each rule ($r_{id}$) the algorithm fills the set of usage decisions $D$ (lines 4-16). To this end, first, it extracts the constraints written in the policy that are associated with each of the specific rules extracted based on their id ($r_{id}$) (line 5). The algorithm then translates these constraints into a policy written in an \textit{interpreted language} and evaluates it (lines 6-7). In the event that the evaluation returns a value $d$ that is true, the algorithm extracts the action associated with it; otherwise it outputs an empty $Usage$ $Decision$ $D$ (lines 8).

In the case where $d$ is true and once the action $a$ is extracted, the algorithm checks if this action is supported by the \textit{interpreted language} by means of the $supported$ procedure (line 10). If the action is eligible for enforcement, it is translated and expressed in the \textit{interpreted language} by means of the procedure $transformAction$ and then enforced with the procedure $perform$. The result of performing the action $V$ is included in the $Usage$ $Decision$ as the tuple $(a,$ $V)$ (lines 11-13). On the contraty, if the action is not eligible for enforcement, the $Usage$ $Decision$ as the tuple $(a,$ $a)$ (line 16), so a third party is informed that the action was not performed by the algorithm either because it cannot be performed as a specific action because it is too abstract or because the third party is responsible for performing it.

\subsection{Running Sample}

In order to provide a running sample using \cref{alg:enforcement}, let us assume two implementations that use Python and SPARQL as \textit{interpreted language}, respectively, and Freemarker as \textit{templated language}. Table~\ref{tab:running} shows a glance of how \cref{alg:enforcement} would enforce the policies of \cref{listing:levela}, \cref{listing:levelb1}, \cref{listing:levelb2}, and \cref{listing:levelc}. For this running sample, it is also assumed as input of \cref{alg:enforcement} the values $\{ ``requestToken": ``eyJhbGciO..."\}$ for the set $M$ and the set $F$ containing the function $request$ implemented with a Java method (since \cref{listing:levelc} is expressed with Freemarker that supports Java-coded functions as \textit{ coded language}). 

Table~\ref{tab:running} presents three columns. The first, common to both implementations, is named $C_R: reduced$ $constraints$ and displays the result running the filtering procedure $constraints$ for each policy; i.e., the line 5 from \cref{alg:enforcement}. It is important to mention that before running the $constraints$ procedure the algorithm has also run procedures $reduce$ and $rules$. The result of the $constraints$ procedure is the tuple $(o, o_{L}, o_{R})$ where $o$ is an operator, $o_L$ and $o_R$ are its left and right operands, respectively. It is worth mentioning that the left operand of \cref{listing:levelb1} had its $[=requestToken]$ Freemarker variable replaced by the value from $M$ under the same name, the left operand of \cref{listing:levelb2} had its $[=timeVar]$ Freemarker variable replaced with the current time injected by the Freemarker engine, and, finally, \cref{listing:levelc} had its $[=tmp]$ variable replaced with the value provided by the API which was retrieved with the $request$ function implemented with Java and passed through the set $F$ to the \cref{alg:enforcement}.

\begin{table*}[h]
\caption{An enforcement running sample using Python and SPARQL as \textit{interpreted languages}}\label{tab:running}
\resizebox{\textwidth}{!}{%
\begin{tabular}{clll}
\hline
Policy   & \multicolumn{1}{c}{$C_R$: reduced constraints}                                                                                                                                                                                                         & \multicolumn{1}{c}{$P_I:$ Python Interpreted Policy}                                                                             & \multicolumn{1}{c}{$P_I:$ SPARQL Interpreted Policy} \\ \hline

\begin{tabular}[c]{@{}l@{}}Policy in \\ \cref{listing:levela}\end{tabular} & \begin{tabular}[c]{@{}l@{}}\{(odrl:gt, odrl:dateTime,\\             ``2018-01-01T00:40:30"\textasciicircum{}\textasciicircum{}xsd:dateTime ), \\  (odrl:eq, odrl:media, \\             ``online"\textasciicircum{}\textasciicircum{}xsd:string ) \}\end{tabular} & \begin{tabular}[c]{@{}l@{}}odrl\_gt(odrl\_datetime(), \xspace cast\_dateTime(``2018-01-01T00:40:30")) \\ \xspace \xspace and odrl\_eq(odrl\_media(), cast\_string(``online"))\end{tabular} & \begin{tabular}[c]{@{}l@{}}SELECT ?ruleId \{ BIND (  \\ \xspace \xspace fnc:now() \textgreater xsd:dateTime(``2018-01-01T00:40:30") AND\\         \xspace \xspace  fnc:media() == xsd:string(``online")\\      AS ?ruleId) \}\end{tabular} \\
                                     
\begin{tabular}[c]{@{}l@{}}Policy in \\ \cref{listing:levelb1}\end{tabular} & \begin{tabular}[c]{@{}l@{}}\{ (eq, ``eyJhbGciO..."\textasciicircum{}\textasciicircum{}xsd:string, \\             ``eyJhbGciO..."\textasciicircum{}\textasciicircum{}xsd:string)\}\end{tabular}                                                    &  \begin{tabular}[c]{@{}l@{}} odrl\_eq(cast\_string(``eyJhbGciO..."), \\ \xspace  \xspace  cast\_string(``eyJhbGciO...")   \}\end{tabular}                                                                                 & \begin{tabular}[c]{@{}l@{}}SELECT ?ruleId \{ BIND (  \\          \xspace \xspace xsd:string(``eyJhbGciO") == xsd:string(``eyJhbGciO...")\\      AS ?ruleId) \}\end{tabular}                                                   \\
\begin{tabular}[c]{@{}l@{}}Policy in \\ \cref{listing:levelb2}\end{tabular} & \begin{tabular}[c]{@{}l@{}}\{ (gt,``13:30"\textasciicircum{}\textasciicircum{}xsd:time, \\             ``08:00:00"\textasciicircum{}\textasciicircum{}xsd:time) \}\end{tabular}                                                                   & odrl\_gt(cast\_time(``13:30"), \xspace  cast\_time(``08:00:00"))    & \begin{tabular}[c]{@{}l@{}}SELECT ?ruleId \{ BIND (  \\    \xspace \xspace        xsd:time(``13:30") \textgreater \xspace  xsd:time(``08:00:00")\\      AS ?ruleId) \}\end{tabular}                                                    \\

\begin{tabular}[c]{@{}l@{}}Policy in \\ \cref{listing:levelc}\end{tabular} & \begin{tabular}[c]{@{}l@{}}\{ (gt,``32"\textasciicircum{}\textasciicircum{}xsd:float, \\             ``35"\textasciicircum{}\textasciicircum{}xsd:float) \}\end{tabular}                                                                          & odrl\_gt(cast\_float(``32"), \xspace  cast\_float(``35"))                                                                                      & \begin{tabular}[c]{@{}l@{}}SELECT ?ruleId \{ BIND (  \\   \xspace \xspace        xsd:float(``32") \textgreater \xspace  xsd:float(``35")\\      AS ?ruleId) \}\end{tabular}                       \\ \hline
\end{tabular}%
}
\end{table*}

The second and third columns of Table~\ref{tab:running} present the result of running the $transform\-Cons\-traints$ procedure on line 6 of \cref{alg:enforcement} for each policy shown in Listings 1 to 4. The result is an equivalent policy $P_I$ expressed according to an \textit{interpreted language}; in the case of the second column is Python and in the case of the third column is SPARQL.  

For Python, it can be observed that a possible implementation of the procedures $transformConstraints$ and $transformAction$ may rewrite each of the reduced constraints $C_R$ as a set of binary functions where the name of the function is the operator and as an argument the operands; then these constraints are concatenated with the Python Logical Operator $and$. Note that each operand, left or right, is casted into a Python type that is equivalent to their \textit{xsd} type. This step is crucial since certain operands act differently depending on the types; e.g., comparing two dates is not the same as comparing dates with time. 

For implementing the $evaluate$ procedure, the \textit{eval} Python function~\footnote{\url{https://www.w3schools.com/python/ref_func_eval.asp}} may be used. Note that, for enforcing policies from Listings 1 to 4, it is also crucial to have implemented in Python the functions that appear in the \textit{Python Interpreted Policy} column, such as \textit{odrl\_datetime} or \textit{cast\_datetime}. The \textit{eval} native Python function will try to invoke these functions when evaluating the policies from \textit{Python Interpreted Policies}.

For SPARQL, it can be observed that a possible implementation of the $transform\-Cons\-traints$ procedure may wrap each of the reduced constraints $C_R$ as a SELECT query with one BIND statement; which would concatenate every constraint with the AND operator from SPARQL. The same approach can be taken for the implementation of $transformAction$. Note that the constraints are rewritten as custom binary SPARQL functions, which is why their prefix changes to \textit{fnc}. Also, note that the \textit{xsd} types are preserved for the same casting reasons as mentioned above. For implementing the $evalue$ procedure, any SPARQL library can be used insofar as it supports extension of the language with native functions; which is needed to implement non-SPARQL native functions \textit{fnc:now} and \textit{fnc:media}. Note that SPARQL needs less custom functions, as it already supports most of the operations needed. For example, the native operand $>$, or \textit{==}, already support dates or numbers as arguments.  

\subsection{Discussion: limitations and challenges ahead}

Once the approach on how to express the ODRL policies to enforce them and handle dynamic data is presented (Section~\ref{sec:classification}) and \cref{alg:enforcement}, it is important to analyse and highlight several limitations: about the ODRL vocabulary and how it affects \cref{alg:enforcement} and about our proposal. In addition, future challenges are discussed.

\textbf{ODRE bad practices:} The approach presented in the article presents a drawback that may imply having different ODRL policies that may seem different but are the same. In addition, these policies would not clearly codify the constraints under which a resources can be used. This is due to the fact that certain policies may mix RDF and an \textit{interpolated} or \textit{templated language} which may also inject some behaviour using a \textit{coded language}. Specifying these constraints without the semantics provided by the ODRL ontology is possible but is a bad practice since, on the one hand, looses the goal of using an ontology that provides a unequivocally definition of a concept, and, on the other hand, makes policies harder to compare, understand by a person, or to write since they will enclose a large amount of behaviour with languages different from the descriptive. 

\textbf{Synchronous vs asynchronous enforcement:} the enforcement of a policy may happen synchronously; whether someone desires to access a certain resource and then the enforcement procedure is invoked. Alternatively, enforcement may occur asynchronously, the constraints of a policy are continuously evaluated and when they enforce positively their related action is performed. Note that these two methods are not exclusive and, instead, there should be policies that must be enforced synchronously and others asynchronously. To this end, ODRL lacks of the semantics to label this feature in a policy and it will be a future challenge to tackle. Also, the \cref{alg:enforcement} from the ODRE framework has been endowed and tested only to perform synchronous enforcement.

\textbf{ODRL abstract definitions:} the ODRL vocabulary presents some important limitations in terms of expressiveness for its operands, operators, and actions. Some of them have abstract  definition and, therefore, it is almost impossible to implement them or, several implementations may drastically differ one from the other depending on the particular use case. For instance, the action $read$ can be interpreted as gaining the access to a certain API or as if the result of the enforcement is the output of reading such API; other interpretations are also possible, as if someone can read in the physically world a document. This issue can be handled using ODRE by relying on the \textit{interpolated/templated language} however, this also entails losing certain semantics as already explained before. This can be view as a benefit that ODRE offers, but also, as a bad practice.

The only approach to solve this issue correctly, without losing semantics or incurring in bad practices, is to extend the ODRL ontology to such specific use cases, providing operands, operators, and actions that are unequivocally described and, therefore, can only be implemented and interpreted in a unique way.

\textbf{ODRL binary predicates:} the ODRL ontology defines the constraints as an element that relates to two operands and an operator. In other words, the operators are terms designed to consider only two inputs, i.e., the left and right operands. This may be a limitation in the case other operators are needed and, these, require more operators than two. For instance, a policy that grants usage if two WGS84 coordinates are located in a proximity threshold of less than 1 Km. To express such policy it will be necessary to define an operator that considers the two coordinates and the threshold of proximity, i.e., three input arguments. 

To tackle this issue, it would seem as a good idea to define a new operator that supports a third operand; however, how to name this new operand is a challenge itself, since the current semantics for operands are left and right. If we scale the input of operators to have N-ary predicates, this problem becomes even more complex. In addition, operands hold no order for the operator explicitly; which is also a potential problem particularly for N-ary predicates. As before, this issue can be addressed using ODRE and relying on the \textit{interpolated/templated language}, however, although it may partially solve this problem, it will also entail the bad practice explained in the first place.

\textbf{ODRL 0-ary operands and actions:} the ODRL standard presents operands as constant values or as ontology terms that must be interpreted and replaced during the enforcement with a value. Nevertheless, the current representation of these operands is limited since they assume that such operands require no input to produce an output value. For instance, \textit{odrl:dateTime} requires no input to provide the current date with time. Far from practical scenarios, some operands may require input values to operate correctly and this, as it is now endowed the ODRL ontology, can not be specified. For instance, \textit{odrl:dateTime} may require as input the timezone to provide the correct date and time. Similar to operands, ODRL standard presents actions as ontology terms that must be interpreted. However, these actions may also require some input information to be performed. These kind of scenarios are currently out of the scope of ODRL but, as it has been adopted in wide spread scenarios like Data Spaces, is a challenge ahead. In addition to this, a challenge ahead is to consider having operands that have as input nested operands, or actions that take as input some operand. As before, this issue can be addressed using ODRE and relying on the \textit{interpolated/templated language}, however, although it may partially solve this problem, it will also entail the bad practice explained in the first place.

\textbf{ODRL Stone-written conditions:} the ODRL ontology relies on terms that are defined as left or right operands to specify data that is not specifically written in the policy but somehow must be used or considered when enforcing. However, the information output as result of interpreting these terms is always compared to data conditions, usually right operands, that are constant. For instance, the policy from Listing~\ref{listing:levela} compares the $dateTime$ left operand with the constant \textit{2018-01-01}. Note that the latter value is stone written in the policy and readable if the policy is accessed, leading to a potential privacy leak. Although this issue may seem not important, if we have a look at the policy from Listing~\ref{listing:levelb1} one can soon realise how leaking a token API is a severe security breach. 

In addition, in the ODRL ontology, there is an unclear specification of where the data come from. For example, with the $dateTime$ left operand it can be assumed that is the date and time of the system that runs the enforcement algorithm. However, the policy shown in the Listing~\ref{listing:levelb1} clearly needs the input of a person, that is, the token, and should rely on a third-party API to check if the token exists instead of having the stone-written value as the right operand. In other words, policies may require data from the system that runs the enforcement algorithm, data from the person trying to access a resource, and data coming from a third party entity (service or person). However, this issue is not addressed by the ODRL standard or their ontology, which is an important challenge. The ODRE approach provides technical support for this problem by relying on the \textit{interpolated, templated} or $coded$ $languages$. Note that in these cases, the bad practice explained in the first place is not happening since no semantic is bypassed with terms outside the ontology.

\textbf{ODRL Responsibilities:} since the ODRL standard has only promoted the ODRL ontology and the enforcement is currently out of the scope of this standard, there is lack expressiveness about the responsibilities related to a policy. For instance, who must perform an action written in the policy? is the enforcement algorithm or an external actor?; who is the responsible of checking the constraints of a policy?, a third-party actor (that may be an enforcement algorithm), a local enforcement algorithm, or a shared task?; which actors are eligible to try to use a resource and thus trigger the enforcement algorithm?. Considering the operand $odrl:spatial$ that stands for geospatial named area who is the responsible of providing such data? There is an unclear answer to all these questions and how these responsibilities should be specified. 

\section{Evaluation}\label{sec:experiments}

In order to evaluate \cref{alg:enforcement} and prove that the ODRE approach is language agnostic, which means that different languages can be used as \textit{interpreted, interpolated, templated,} or \textit{coded}, two implementations are provided: $pyodre$ and $odre{-}java$. Both implementations are published on GitHub and distributed as open-source under the APACHE 2.0 licence. $Pyodre$\footnote{\url{https://github.com/ODRE-Framework/odre-python}} relies on Python as \textit{interpreted language}, Jinja2 as \textit{interpolated} and \textit{templated language}, and Python as \textit{coded language}. This implementation is also distributed through the $pip$ package installer\footnote{\url{https://pypi.org/project/pyodre/}}. The $odre{-}java$\footnote{\url{https://github.com/ODRE-Framework/odre-java}} relies on SPARQL as \textit{interpreted language}, Freemarker as \textit{interpolated and templated language}, and Java as \textit{coded language}. This implementation is distributed through maven central\footnote{\url{https://central.sonatype.com/artifact/io.github.odre-framework/odre-core}}.

Both implementations cover most of the operands from ODRL, namely: \textit{odrl:lt}, \textit{odrl:lteq}, \textit{odrl:eq}, \textit{odrl:neq}, \textit{odrl:gt}, \textit{odrl:gteq}. Also, they implement the left operand \textit{odrl:date\-Time}. In addition, to prove the extensibility of our proposal, a small extension of the ODRL vocabulary about time has been created. The ontology of this extension is available at \url{https://w3id.org/def/odre-time#}, let us assume $otime$ as its prefix. The ontology extension includes a new operand named $otime{:}time$ that provides the current time, a new operator $otime{:}between$ that returns true if the current time is between the ones specified as the left and right operands. Finally, as another extension, the action $dummy{:}read$ has been provided to test the enforcement of the actions. This action reads a dummy API and provides as output its result. Note that this action has no ontology related, since it has been created only to show how actions can be enforced but is not meant to be used in real scenarios. 

In order to test these implementations, 24 policies have been defined using the following criterion: 6 policies defined as the left and right operands a constant and each one tested one of the operators $lt$, $lteq$, $eq$, $neq$, $gt$, $gteq$; 6 policies defined with the left operand $odrl{:}dateTime$ and each tested one of the previous operators; 6 policies defined with the left operand $otime{:}time$ and each tested one of the previous operators; 1 policy defined two constants and tested the $otime{:}between$ operand; 1 policy defined two constants and tested the $dummy{:}read$; and, finally, 4 policies used the their \textit{templated language} to inject values or functions (implemented with Python or Java depending on the implementation). Then, these policies were used to define 24 unit tests in both Python\footnote{\url{https://github.com/ODRE-Framework/odre-python/tree/main/test}} and Java\footnote{\url{https://github.com/ODRE-Framework/odre-java/blob/main/src/test/java/tests/odre}} to test their respective implementations.

\begin{figure}[h]
\includegraphics[width=\textwidth]{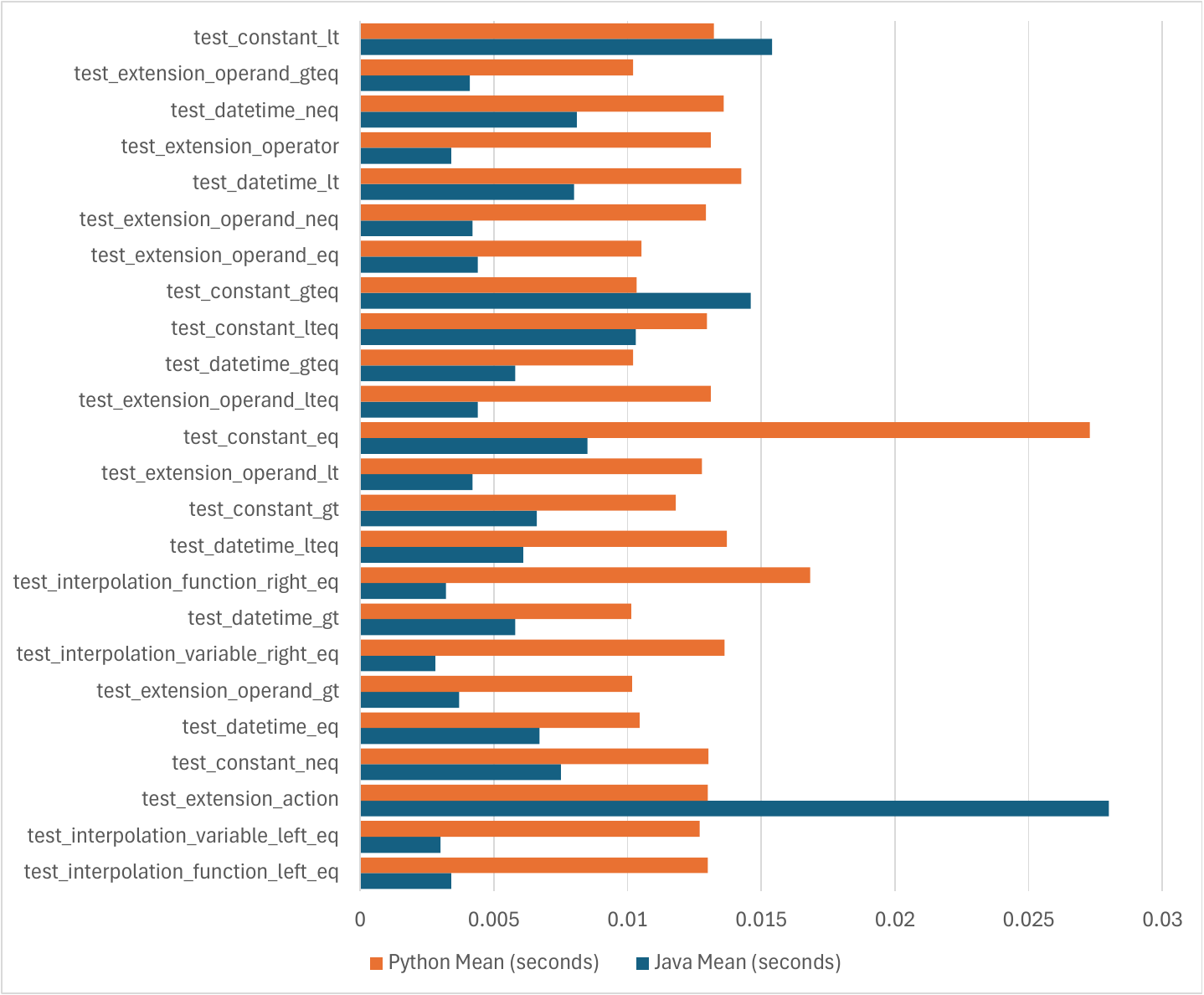}
\centering
\caption{ODRL enforcement duration in seconds for Python and Java implementations}
\label{fig:results}
\end{figure}

In addition to running the unit tests successfully, the time it takes to carry out the enforcement has been analysed. To this end, each unitary test was run 15 times and the seconds they took recorded. The first five measurements were discarded as warm-up, and the rest were averaged and compared between implementations. Figure~\ref{fig:results} displays these results, showing the results obtained by the Python and Java implementations to run the same policy in seconds. 

It can be observed that, in general, both implementations enforce all the policing in less than 0.03 seconds. It is worth mentioning that in general, the Java implementation behaves better than python with the exception of $test\_extension\_action$, the difference observed can be due to the fact that such policy performed a request to an API and Java handles this functionality with streams, increasing the running time. For the sake of reproducibility, and following the FAIR principles, a frozen version of the implementations have been uploaded to the Zenodo repository\footnote{\url{https://zenodo.org/records/13735971}}, along with two reports about the implementations and instructions for each on how reproduce these experiments. 

\section{Related Work}\label{sec:relatedwork}

The definition of policies for the usage of data resources has been extensively analysed by researchers~\cite{DBLP:journals/computing/AlamoGGD22}. However, it is important to differentiate between two types of research proposals; those focussing on the description of these policies and related challenges, and those focussing on the enforcement of these policies~\cite{8962144}. In addition, another feature to take into account when analysing proposals similar to ODRE is the fact that they rely, or not, on standards vocabularies to express the policies. In this article, the proposals are narrowed down to those that rely on policies expressed with the W3C ODRL standard and which scope is the enforcement of ODRL policies, discarding the rest as out of scope. The proposals related to enforcement are numerous~\cite{akaichi2022usage,DBLP:journals/semweb/KirraneVd18}, in order to ease their review, the authors have classified them into the following types:

\textbf{Theoretical enforcement {-}} these are proposals that promote theoretical ideas to perform the enforcement of ODRL policies, or challenges, but without implementations to support their claims. The proposal of Munoz-Arcentales et al.~\cite{su12093885} presents the need for flexible usage control solutions that can be adapted and used in different scenarios, but enforcing ODRL is out of the scope. The proposal of Cirillo et al.~\cite{9314823} rely on a subset of ODRL and does not tackle how to enforce the ODRL policies, but instead, policies expressed with an extended version of ODRL for specific use cases. The proposal of Cirillo et al.~\cite{9314823} is not considered to enforce ODRL since it does not support policies expressed only with ODRL terms. The proposal of Akaichi et al.~\cite{akaichi2024interoperable} presents an architecture to handle ODRL policies in scenarios that require dynamic constraints (addressing the limitation \textit{ODRL Stone-written conditions}). Their proposal focuses on a specific client-server scenario and do not delve in the details of how to express the ODRL policies from a generic point of view of the enforcement. In addition, their proposal extends the ODRL ontology in such a way that their solution works only with it. Finally, their proposal is theoretical and has no implementation related.

\textit{Discussion:} some of these proposals ~\cite{su12093885, 9314823} do not actually aim to support ODRL enforcement but, instead, use a subset of the ODRL for certain limited scenarios not promoting a general purpose enforcement solution. Furthermore, these two proposals do not provide a generic algorithm for enforcing ODRL. The proposal closer to ODRE is the one of Akaichi et al.~\cite{akaichi2024interoperable} since they addressed one of the limitations that also ODRE tackles. However their proposal is not generic for the ODRL ontology or any extensions of it. In addition, they provide no generic algorithm for enforcing ODRL.

\textbf{Practical enforcement {-}} these are proposals available on the Web. ODRL-PAP\footnote{\url{https://github.com/wistefan/odrl-pap}} promotes an enforcement engine in the context of the DOME-Marketplace project\footnote{\url{https://dome-marketplace.eu}}. The MOSAICroWN Policy Engine\footnote{\url{https://github.com/mosaicrown/policy-engine}} is built in the context of the MOSAICrOWN project\footnote{\url{https://mosaicrown.eu}}. Both of the previous proposals are a specific solutions for projects and have their own vocabulary based on ODRL, but different from the ODRL standard vocabulary. 

\textit{Discussion:} Although these proposals perform the enforcement, the policies they support is limited and are not ODRL policies expressed with the standard ontology. Furthermore, as major drawback these proposals are not related to research papers and do not provide any procedure pseudo-code to reproduce their results. 

\textbf{Traverse enforcing {-}} these are proposals focussing on the enforcement of ODRL policies by traversing the policies into another vocabulary to express policies that support enforcement~\cite{akaichi2022usage}. Some proposals focus on aligning ODRL with the XACML standard~\cite{standard2013extensible} and its architecture~\cite{dam2023policy}. However, although these proposals may benefit from the enforcement of XACML, it is not possible to fully align ODRL with XACML and, therefore, this approach can be suitable for certain scenarios but not as a general solution to enforce ODRL; which is the goal of our article. Another similar approach was proposed by Hosseinzadeh et al.~\cite{hosseinzadeh2020systematic}; which consists of translating ODRL policies into MYDATA Policy Language\footnote{\url{https://developer.mydata-control.de/language/}}. Another approach was proposed in a position article that opened the possibility of enforcing ODRL policies as mappings~\cite{10.1145/3543873.3587358}. These are commonly used to translate heterogeneous data sources into RDF~\cite{manola2004rdf}, i.e., the format of the ODRL policies. Again, this proposal is suitable for certain scenarios but has major drawbacks, the most important one is that the policies must always be expressed mixing the terms from ODRL and the mapping language. 

\textit{Discussion:} These are the proposals closer to the one presented in this article, however, they do not provide a general algorithm to enforce ODRL using the standard vocabulary, or extensions of it, as presented in Section~\ref{sec:classification}. These proposals rely on translating ODRL policies into others expressed with vocabularies that have enforcing capabilities. In other words, these proposals rely on translating $\emptyset{-}Level$ policies expressed with the ODRL ontology into a different, yet somehow equivalent, $\emptyset{-}Level$ policy expressed with another vocabulary. This approach has to main drawbacks, on the one hand, the ODRL vocabulary is not fully equivalent to others (and vice versa) and thus not any ODRL policy can be translated, on the other hand, in the case of extending ODRL it will be necessary to extend also the vocabulary to which the policies will be translated and their enforcement algorithms. As a result, these approaches are suitable for certain scenarios but do not provide a genuine enforcement procedure for ODRL and, instead, rely on the enforcement of other proposals.

The approach of using a metalanguage to enforce ODRL policies presented in our previous article~\cite{10.1145/3543873.3587358} was the basis to build the ODRE framework. However, the approach presented by Cano-Benito et al.~\cite{10.1145/3543873.3587358} only considered the policies $B1{-}Level$ or above policies and did not provide the analysis presented in Section~\ref{sec:classification}. In addition, no algorithm was provided.

To the best of the author's knowledge, this article is the first presenting a generic enforcement procedure with an algorithm for ODRL policies that supports the standard and provides several extension mechanisms. In addition, in the articles reviewed in this section, the different key points discussed in the section~\ref{sec:conclusions} have not been tackled by the authors. In conclusion, the ODRE framework presents a novel approach to support the ODRL standard and allow ODRL enforcement; which is a feature demanded for this standard~\cite{akaichi2022usage}.

\section{Conclusions}\label{sec:conclusions}

In this article, the ODRE framework has been presented including: a novel approach to write and classify ODRL policies enabling their enforcement, a generic enforcement algorithm, and two implementations. In addition, an analysis on the current limitations of ODRE and ODRL has been presented along with future challenges. Previous proposals have delved into how to enforce ODRL: some exclusively from the theoretical point of view, others providing only implementations without research articles, and others traversing the enforcement to other initiatives. These latest proposals have mainly relied on translating ODRL policies into other vocabularies that already supported enforcement (such as XACML). However, these proposals fall short when used to enforce any ODRL policy, since the expressiveness (operands, operators, actions, and more) present in the ODRL ontology does not always exist in that target vocabulary. To the best of our knowledge, ODRE is the first proposal that provides a solution to enforce ODRL and is flexible enough to support future extensions of ODRL for specific domain challenges. In addition, the experiments carried out prove that the implementations perform enforcement efficiently for real-world scenarios.

In the future, the authors will address the limitations of ODRL and try to tackle challenges identified; such as defining N-ary operators or operands or addressing how to express synchronous and asynchronous enforcement in policies and extend ODRE to support the second. In addition, specific extensions will be developed for the time and geospatial domains. Regarding ODRE limitations, the authors plan to publish good-practices and guidelines documentation to prevent bad practices. Finally, the authors will analyse the impact on the enforcement that may produce using ODRL with other related vocabularies, preferably standard, such as the Data Privacy Vocabulary\footnote{\url{https://w3c.github.io/dpv/dpv/}} (DPV) to enrich the expressiveness of usage and privacy use cases.

\section*{Acknowledgments}
This work is partially funded by the European Union’s Horizon 2020 Research and Innovation Programme through the AURORAL project, Grant Agreement No. 101016854.

\bibliographystyle{elsarticle-num} 
\bibliography{bibliography}

\begin{thebibliography}{10}
\expandafter\ifx\csname url\endcsname\relax
  \def\url#1{\texttt{#1}}\fi
\expandafter\ifx\csname urlprefix\endcsname\relax\def\urlprefix{URL }\fi
\expandafter\ifx\csname href\endcsname\relax
  \def\href#1#2{#2} \def\path#1{#1}\fi

\bibitem{10.1145/2295136.2295159}
F.~Kelbert, A.~Pretschner, \href{https://doi.org/10.1145/2295136.2295159}{Towards a policy enforcement infrastructure for distributed usage control}, in: Proceedings of the 17th ACM Symposium on Access Control Models and Technologies, SACMAT '12, Association for Computing Machinery, New York, NY, USA, 2012, p. 119–122.
\newblock \href {https://doi.org/10.1145/2295136.2295159} {\path{doi:10.1145/2295136.2295159}}.
\newline\urlprefix\url{https://doi.org/10.1145/2295136.2295159}

\bibitem{6984194}
M.~Henze, L.~Hermerschmidt, D.~Kerpen, R.~Häußling, B.~Rumpe, K.~Wehrle, User-driven privacy enforcement for cloud-based services in the internet of things, in: 2014 International Conference on Future Internet of Things and Cloud, 2014, pp. 191--196.
\newblock \href {https://doi.org/10.1109/FiCloud.2014.38} {\path{doi:10.1109/FiCloud.2014.38}}.

\bibitem{akaichi2024interoperable}
I.~Akaichi, W.~Slabbinck, J.~A. Rojas, C.~Van~Gheluwe, G.~Bozzi, P.~Colpaert, R.~Verborgh, S.~Kirrane, Interoperable and continuous usage control enforcement in dataspaces, in: The Second International Workshop on Semantics in Dataspaces, co-located with the Extended Semantic Web Conference, 2024.

\bibitem{standard2013extensible}
O.~Standard, extensible access control markup language (xacml) version 3.0, A:(22 January 2013). URl: http://docs. oasis-open. org/xacml/3.0/xacml-3.0-core-spec-os-en. html (2013).

\bibitem{odrl-standard}
R.~I. Monegraph, S.~Villata, {ODRL Information Model 2.2} (2018).

\bibitem{10.5555/2874359.2874375}
V.~Rodr\'{\i}guez-Doncel, A.~G\'{o}mez-P\'{e}rez, N.~Mihindukulasooriya, Rights declaration in linked data, in: Proceedings of the Fourth International Conference on Consuming Linked Data - Volume 1034, COLD'13, CEUR-WS.org, Aachen, DEU, 2013, p. 158–169.

\bibitem{de2019odrl}
M.~De~Vos, S.~Kirrane, J.~Padget, K.~Satoh, Odrl policy modelling and compliance checking, in: Rules and Reasoning: Third International Joint Conference, RuleML+ RR 2019, Bolzano, Italy, September 16--19, 2019, Proceedings 3, Springer, 2019, pp. 36--51.

\bibitem{10.1145/2660517.2660530}
S.~Steyskal, A.~Polleres, \href{https://doi.org/10.1145/2660517.2660530}{Defining expressive access policies for linked data using the odrl ontology 2.0}, in: Proceedings of the 10th International Conference on Semantic Systems, SEM '14, Association for Computing Machinery, New York, NY, USA, 2014, p. 20–23.
\newblock \href {https://doi.org/10.1145/2660517.2660530} {\path{doi:10.1145/2660517.2660530}}.
\newline\urlprefix\url{https://doi.org/10.1145/2660517.2660530}

\bibitem{eitel2021usage}
A.~Eitel, C.~Jung, R.~Brandst{\"a}dter, A.~Hosseinzadeh, S.~Bader, C.~K{\"u}hnle, P.~Birnstill, G.~Brost, M.~Gall, F.~Bruckner, et~al., Usage control in the international data spaces, Aufl. IDS Association, Berlin (2021).

\bibitem{10.1007/978-3-030-89811-3_4}
M.~G. Kebede, G.~Sileno, T.~Van~Engers, A critical reflection on odrl, in: V.~Rodr{\'i}guez-Doncel, M.~Palmirani, M.~Araszkiewicz, P.~Casanovas, U.~Pagallo, G.~Sartor (Eds.), AI Approaches to the Complexity of Legal Systems XI-XII, Springer International Publishing, Cham, 2021, pp. 48--61.

\bibitem{akaichi2022usage}
I.~Akaichi, S.~Kirrane, Usage control specification, enforcement, and robustness: A survey (2022).
\newblock \href {http://arxiv.org/abs/2203.04800} {\path{arXiv:2203.04800}}.

\bibitem{chomsky2014aspects}
N.~Chomsky, Aspects of the Theory of Syntax, no.~11, MIT press, 2014.

\bibitem{parr2004enforcing}
T.~J. Parr, Enforcing strict model-view separation in template engines, in: Proceedings of the 13th international conference on World Wide Web, 2004, pp. 224--233.

\bibitem{manola2004rdf}
F.~Manola, E.~Miller, B.~McBride, et~al., Rdf primer, W3C recommendation 10~(1-107) (2004) 6.

\bibitem{sparql-standard}
S.~Harris, A.~Seaborne, {SPARQL 1.1 Query Language} (2013).

\bibitem{suarez2015neon}
M.~C. Su{\'a}rez-Figueroa, A.~G{\'o}mez-P{\'e}rez, M.~Fernandez-Lopez, The neon methodology framework: A scenario-based methodology for ontology development, Applied ontology 10~(2) (2015) 107--145.

\bibitem{DBLP:journals/computing/AlamoGGD22}
J.~M. del {\'{A}}lamo, D.~S. Guam{\'{a}}n, B.~Garc{\'{\i}}a, A.~Diez, \href{https://doi.org/10.1007/s00607-022-01076-3}{A systematic mapping study on automated analysis of privacy policies}, Computing 104~(9) (2022) 2053--2076.
\newblock \href {https://doi.org/10.1007/s00607-022-01076-3} {\path{doi:10.1007/s00607-022-01076-3}}.
\newline\urlprefix\url{https://doi.org/10.1007/s00607-022-01076-3}

\bibitem{8962144}
J.~Leicht, M.~Heisel, A survey on privacy policy languages: Expressiveness concerning data protection regulations, in: 2019 12th CMI Conference on Cybersecurity and Privacy (CMI), 2019, pp. 1--6.
\newblock \href {https://doi.org/10.1109/CMI48017.2019.8962144} {\path{doi:10.1109/CMI48017.2019.8962144}}.

\bibitem{DBLP:journals/semweb/KirraneVd18}
S.~Kirrane, S.~Villata, M.~d'Aquin, \href{https://doi.org/10.3233/SW-180289}{Privacy, security and policies: {A} review of problems and solutions with semantic web technologies}, Semantic Web 9~(2) (2018) 153--161.
\newblock \href {https://doi.org/10.3233/SW-180289} {\path{doi:10.3233/SW-180289}}.
\newline\urlprefix\url{https://doi.org/10.3233/SW-180289}

\bibitem{su12093885}
A.~Munoz-Arcentales, S.~López-Pernas, A.~Pozo, A.~Alonso, J.~Salvachúa, G.~Huecas, \href{https://www.mdpi.com/2071-1050/12/9/3885}{Data usage and access control in industrial data spaces: Implementation using fiware}, Sustainability 12~(9) (2020).
\newblock \href {https://doi.org/10.3390/su12093885} {\path{doi:10.3390/su12093885}}.
\newline\urlprefix\url{https://www.mdpi.com/2071-1050/12/9/3885}

\bibitem{9314823}
F.~Cirillo, B.~Cheng, R.~Porcellana, M.~Russo, G.~Solmaz, H.~Sakamoto, S.~P. Romano, Intentkeeper: Intent-oriented data usage control for federated data analytics, in: 2020 IEEE 45th Conference on Local Computer Networks (LCN), 2020, pp. 204--215.
\newblock \href {https://doi.org/10.1109/LCN48667.2020.9314823} {\path{doi:10.1109/LCN48667.2020.9314823}}.

\bibitem{dam2023policy}
T.~Dam, A.~Krimbacher, S.~Neumaier, Policy patterns for usage control in data spaces (2023).
\newblock \href {http://arxiv.org/abs/2309.11289} {\path{arXiv:2309.11289}}.

\bibitem{hosseinzadeh2020systematic}
A.~Hosseinzadeh, A.~Eitel, C.~Jung, A systematic approach toward extracting technically enforceable policies from data usage control requirements., in: ICISSP, 2020, pp. 397--405.

\bibitem{10.1145/3543873.3587358}
J.~Cano-Benito, A.~Cimmino, R.~Garc\'{\i}a-Castro, \href{https://doi.org/10.1145/3543873.3587358}{Injecting data into odrl privacy policies dynamically with rdf mappings}, in: Companion Proceedings of the ACM Web Conference 2023, WWW '23 Companion, Association for Computing Machinery, New York, NY, USA, 2023, p. 246–249.
\newblock \href {https://doi.org/10.1145/3543873.3587358} {\path{doi:10.1145/3543873.3587358}}.
\newline\urlprefix\url{https://doi.org/10.1145/3543873.3587358}

\end{thebibliography}



\end{document}